# Design and analysis of a multi-agent e-learning system using prometheus design tool

**Kennedy E. Ehimwenma, Sujatha krishnamoorthy**
Department of Computer Science, College of Science and Technology, Wenzhou-Kean University, China

| Article Info | ABSTRACT |
|---|---|
| *Article history:*<br><br>Received Apr 5, 2020<br>Revised Dec 31, 2020<br>Accepted Jan 6, 2021<br><br>*Keywords:*<br><br>Agent methodology<br>Computational education<br>First-order logic<br>Pre-assessment classification<br>Requirements engineering | Agent unified modeling languages (AUML) are agent-oriented approaches that supports the specification, design, visualization and documentation of an agent-based system. This paper presents the use of prometheus AUML approach for the modeling of a pre-assessment system of five interactive agents. The pre-assessment system, as previously reported, is a multi-agent-based e-learning system that is developed to support the assessment of prior learning skills in students so as to classify their skills and make recommendation for their learning. This paper discusses the detailed design approach of the system in a step-by-step manner; and domain knowledge abstraction and organization in the system. In addition, the analysis of the data collated and models of prediction for future pre-assessment results are also presented. |

*Corresponding Author:*

Kennedy E. Ehimwenma
Department of Computer Science
Wenzhou-Kean University
88 Daxue Rd, Ouhai District, Wenzhou, Zhejiang, China
Email: kehimwen@kean.edu

## 1. INTRODUCTION

An agent software methodology is a set of guidelines that covers the entire life-cycle of a multi-agent development process. A multi-agent system (MAS) is a system of interactive agents or autonomous program modules. In general, a unified modelling language (UML) assists software developers to specify, design, visualize and document software engineering processes that meets application requirements [1]. A UML allow models to be created, considered, developed, and processed in a standard way from the initial phase of analysis to design and implementation [2]. Systems implementation is focused on users' needs as well as system functionality with requirements specification as the driver. From start to finish, effective and efficient system evolves from user interaction and the incremental principle of development. Software development stages have shared abstraction in both object-oriented programming (OOP) methodology and agent-oriented software engineering (AOSE). In OOP paradigm, these stages are: requirements gathering, analysis, design, implementation, testing and maintenance. Whilst the AOSE process subsumes the steps in OOP methodologies, the concepts for developing objects (in OOP) are however different from those in agent-based systems. For instance, object-oriented methodologies cover concepts such as objects, classes and inheritance. But in AOSE, design concepts are terms that view agents as autonomous, situated, reactive, and social. This paper is a presentation of the application of prometheus [3, 4] agent-oriented methodology for the static and dynamic design of an elearning MAS. Though there are several AOSE methodologies for designing agent-based systems, the choice of prometheus was predicated on its structured and detailed step-by-step procedure that supports how requirement statements can be acquired. The purpose of the system is to pre-assess students' prior learning, classify their skills, and make recommendation for appropriate material suitable to their needs. Thus, the contribution of this paper is: i) To demonstrate requirements analysis and design specifications for the development of an e-learning pre-assessment system using MAS. ii) To analyse the descriptive functions and roles of multi-agents within an e-learning pre-assessment system. iii) To show a detailed model of software engineering with agent UML (AUML) tool for teaching and learning. iv) To demonstrate inter-agent communication for the assessments and classification of students' prior skill-set. v) To analyse the data collated from the system using regression models of prediction. This paper continues with the background logic of knowledge engineering for the system in which an abstract model of an ontology tree traversal is discussed as applicable in the MAS implementation. In section 2, the paper presents AUML tools and agent software development life cycle (ASDLC). In section 3, models of analysis and design from the use of the prometheus design tool (PDT) is presented. Section 4 looks into implementation, issues at experimentation, data collection and analysis; and section 5 is conclusion.

Logic program in decision support system
Logic formulas are formal specifications that are readily used to represent facts, statements and propositions. Such formalism e.g., Horn clauses, answer set programming, first order logic formulas and description logics are used to for reasoning-supported decision processes in a dynamic system. With logic programs and its diverse variants of formalization, facts and objects have been collected, categorized, and relations established in-between objects of facts; and decisions taken e.g., [5, 6]. Approaches for MAS development using first order logic (FOL) have also been demonstrated in [7, 8]. In a hybrid distributed system in which asymptotic consensus result was obtained, [7] presented a leader-follower consensus MAS in which the leader system shared knowledge with the follower system whose description was given in FOL. Similarly, recent studies in MAS e.g. [8-11] have emphasized the need for adaptive elearning systems that can personalize learning so as to meet individual learner needs. This is because what a learner wants, may actually, be different from what he needs to learn. This research addresses this gap in the development of elearning systems.

Background logic of knowledge engineering for the pre-assessment system
Let $p$ be a predicate. A binary relation between objects $x_1$ and $x_2$ can be given symbolically as $p(x_1, x_2)$. Also, let $D$ be a domain of directed ontological nodes, and $i$, $j$, and $k = 1, 2, 3, ..., n, n + 1$, respectively. If $i$ represents the level of hierarchies of nodes on the horizontal traversal, and $j$ the arrangements of nodes on the vertical traversal and $p_{i,j}$ some random predicates or property such that $x_{i,j}$ is a parent node, $x_{i+1,j+1}$ a perquisite parent node next to $x_{i,j}$, and $z_{j,k}$ a leafnode; then the following abstraction holds:

- ✼ $\forall x \in D, p_{i,j}(x_{i,j})$, which states that, every node has a property.
- ✼ $\forall x \in D, p_{i,j}(x_{i,j}, x_{i+1,j+1})$, which means a *parent to prerequisite parent node* relation. Any parent node is a desired topic to be learned by students.
- ✼ $\forall x \in D, p_{i+1,j}(x_{i,j}, z_{j,k})$, that there is a direct relation of a *parent node* to its *own leafnodes*.
- ✼ $\forall x \forall z \in D, p_{i,j}(x_{i,j}, x_{i+1,j+1}) \land p_{i+1,j+1}(x_{i+1,j+1}, z_{j\pm 1,k}) \rightarrow p_{i+1,j+1}(x_{i,j}, z_{j\pm 1,k})$, which is a transitive relation for navigating *leafnodes* connected to *prerequisite parent nodes*, or
- ✼ $\forall x \forall z \in D, p_{i,j}(x_{i,j}, x_{i+1,j+1}) \land p_{i+1,j+1}(x_{i+1,j+1}, z_{j\pm 1,k}) \rightarrow p_{i,j+1}(x_{i,j}, z_{j\pm 1,k})$, which are transitive relations for traversals of *leafnodes* connected to *parent nodes*.

With a tree diagram as shown in Figure 1, the above stated symbolic relations are further deduced as follows: That the parent nodes or the objects of learning are the c1, c2, c3, c4, c5, and c6. Along the horizontal traversal, *Level 1* to *Level 4*, the preceding node e.g., c1 is a parent to c2 and c3; and c2 and c3 are in turn perquisites nodes to c1. Then, c2 and c3 are parent nodes to c4 and c5, respectively; and c4 and c5 are in turn prerequisites to c2 and c3, respectively. Further down, c4 is a parent to c6, and c6 in turn is a prerequisite to c4. In the tree, the set of $c \in C$ have their respective leafnodes $N_i$ labelled as $N_1, N_2, N_3, ..., N_{11}$ are leafnodes $n \in N$.

Establishing navigational relations between nodes
On the basis of the tree as shown in Figure 1, we now show the relationship between nodes in the tree and the symbolic axioms stated earlier as implemented on the multi-agent pre-assessment system that: $p_{i,j}(x_{i,j})$ describes a unary predicate, and an example is *desiredConcept*(c1); and $p_{i,j}(x_{i,j}, x_{i+1,j\pm 1})$ describes a binary relation which states that a parent has a named prerequisite, and an example of the form is *hasPrerequisite*(c1, c2); and $p_{i+1,j}(x_{i+1,j}, z_{j,k})$ which states that a parent has a named leafnode, and an example is formula *hasKB*(c6, $n_{11}$). Then for all pre-assessment and recommendation of any *failed* learning unit i.e., the leafnodes *N*, we state that

$$p_{i,j}(x_{i,j}, x_{i+1,j\pm1}) \wedge p_{i+1,j\pm1}(x_{i+1,j\pm1}, z_{j\pm1,k}) \rightarrow p_{i+1,j\pm1}(x_{i,j}, z_{j\pm1,k}). \qquad axiom\ 1$$

*Axiom 1* states that if a parent node $x_{i,j}$ has a named prerequisite $x_{i+1,j\pm1}$ (one level below the hierarchy) of $x_{i,j}$ either on its right hand or left-hand side which is denoted by $j \pm 1$ (+ for right, and – for left), and the named prerequisite has a named leafnode $z_{j\pm1,k}$, then the parent node has a direct relation with the leafnode like the perquisite has. An example of this transitive closure is *hasPrerequisite(c4, c6) ∧ hasKB(c6, n$_{11}$)* → *hasKB(c4, n$_{11}$)* that satisfies the property of *transitivity*. In addition, the *axiom 1* conveys the leafnodes $z_{j\pm1,k}$ that are: *i)* pre-assessed upon, and *ii)* the nodes that are recommended when any leafnode *N* that is connected to the prerequisite node $x_{i+1,j\pm1}$ are *failed*. On the other hand, the counterpart *axiom 2*

$$p_{i,j}(x_{i,j}, x_{i+1,j\pm1}) \wedge p_{i+1,j\pm1}(x_{i+1,j\pm1}, z_{j\pm1,k}) \rightarrow p_{i+1,j}(x_{i,j}, z_{j\pm1,k}) \qquad axiom\ 2$$

is the *axiom* that also satisfies the property of *transitivity*. In this case, it is for the recommendation of leafnodes $z_{j\pm1,k}$ that has direct relations to the desired topic's $x_{i,j}$ given that an episode of pre-assessment on the perquisite $z_{j\pm1,k}$ connected to $x_{i+1,j\pm1}$ have all been attempted and are all *passed*. An example of this logical axiom 2 is *hasPrerequisite(c1, c3) ∧ hasKB(c3, {n4,n5, n6})* → *hasKB(c1, n1)*.

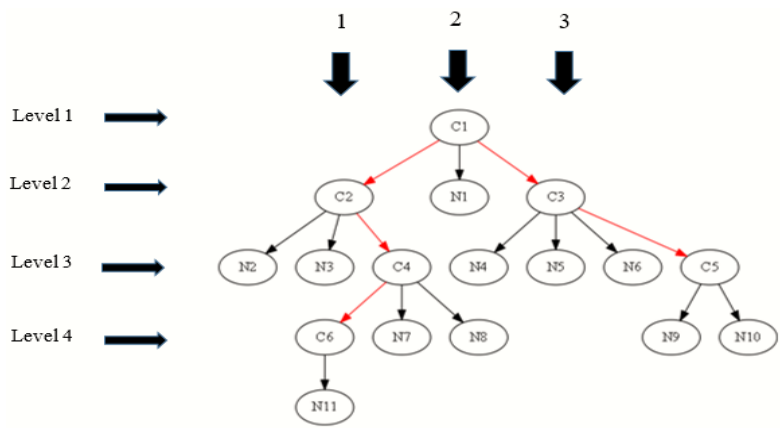

Figure 1. A knowledge graph of multiple horizontal and vertical traversal

In our agent-based pre-assessment system, agents need to communicate the ground fact representation of this logical axioms. For instance, for an agent to resolve the relevant plans for their next action, this group of agents must inter-communicate the *desired topics*, *passed leafnodes* and/or *failed leafnodes* using the following predicate logic form *passed($z_{j+1,k}$)* and *failed($z_{j+1,k}$)*. The predicates which are the actions taken by the multi-agents based on students' response to questions, form the basis of the facts about the outcome of a student using logic programming. This is because any object has a property that it satisfies or that any object is connected by some relation to another object. From the foregoing, the explicitly stated logic-based formulas are the premises on which the multi-agent of the pre-assessment system interacts, select ontological nodes, select questions associated with leafnodes, assess users, classify user skills and recommend learning materials.

## 2. RELATED WORK ON AUML AND E-LEARNING SYSTEMS

A UML and AUML diagrams are architectural model for information systems development. AUML in particular, has been used in the design and specification of systems in the areas of weather forecasting, business trading, petri-net, miner robot; and agent negotiation before implementation. In the field of educational systems, agent-based system development research on intelligent agent models for elearning have received little attention. Amongst this few are the works of [12] and [13]. In [12], Gaia methodology was applied in the analysis and design of an elearning based MAS. The system which was proposed for implementation on JADE framework is a security-based MAS that was meant to detect threats and provide protection on web-based learning management systems (LMS) such as Moodle. In [13] an analysis and design of a web-based services for virtual learning environment (VLE) was also discussed. The paper conceptualized a VLE application on mobile agent technology for the assessment of students' knowledge, and described agent role and agent interaction using a UML tool, and finally to implementation using JADE. The InfoStation system [2] is a project of distributed elearning centre (DeLC), also used multi-agent technology with proposed implementation on JADE [14]. With a UML, [2] described the InfoStation system as a system of interactive agents whose functions included designated e-services. Also, in [15] the AGILE-PASSI methodology was reported as the development tool for a medical educational game called MEDEDUC for the purpose of improving learning in medical education and clinical performance. As a, MEDEDUC allowed students to answer questions at different level of difficulty on multimedia presentation. While many applications on agent-based technology are developed in fields such as commerce and security, or adaptive dynamic programming [16] very limited attention has been given to agent-based development for student learning. Among the aforementioned few, none had the combined system goal of skills classification and recommendation of learning materials that we are presenting in this paper.

### 2.1. Agent-oriented methodology

Methodologies as a process of engineering a software enables developers to concretise the various interaction components of a system and the functions needed amongst the various components for a system to be realised. The work of [17] acknowledged that several AOSE methodologies exists for the analysis and design of MAS but there also exists difficulty in the choice of the appropriate methodology for software solutions from domain to domain. For instance, amongst gaia [18-21], tropos [22, 23], MaSE [24], PASSI [25, 26] and prometheus [27] methodologies what factor should inform the developer's choice? Though these methodologies show similarities in their design process, there are however a varying degree of differences: From requirements analysis, through to functionality modelling for agents, and implementation. In the following section, Promethous is presented; and in Table 1 is a comparative summary of Gaia, Tropos and Prometheus methodologies with regards to their similarities and differences, and the basis upon which the Prometheus methodology was choosing for the design of the pre-assessment system. As shown in Table 1, one common similarity is the customised tool associated with each methodology to support their engineering process. The difference in their respective engineering process can be found in their design steps. For instance, the *tropos* concept of *softgoals* which is equivalent to *subgoals* in prometheus is a breakdown of *hardgoals* and *initial goal* of agents (or actors) functionalities, respectively. This, [28] referred to as *role decomposition* which reduces the complexity in MAS engineering. The basis for prometheus is the design step for deriving *initial goals*.

Table 1. A comparative summary of gaia, tropos and prometheus

| Methodologies | Phases | Comparison |
|---|---|---|
| Gaia | * Statement of requirement<br>* Analysis<br>* Design | * Lack detailed step-by-step breakdown.<br>* No details on how requirement statements may be acquired.<br>* View agent system as an organisational model.<br>* Roles are similar to functionalities in Prometheus.<br>* Editor tool Gaia4E supports design. |
| Tropos | *Early requirement phase<br>* Later requirement phase<br>* Architectural design<br>*Detailed design<br>* Implementation | * Emphasises the *Early Requirement Analysis*, then the *Later Requirement Phase*.<br>* Specialisation of Goals into subclasses of *Hardgoal*, and *Softgoals* for actors of system.<br>* No general architecture containing all the phases of design as in Gaia, MaSE, or Prometheus.<br>* Has a design support tool called Taom4E. |
| Prometheus | * System specification<br>* Architectural design<br>* Detailed design phase | * No Early Requirement phase as in Tropos. But this can be adapted.<br>* Uses *Initial goals*, that are refined or broken down into *Subgoals* for agents.<br>* Very detailed design activity from System Specification phase to other phases.<br>* Reliance on expert knowledge on domain subject for requirement acquisition.<br>* Has a customised PDT, a AUML tool that supports design process. |

### 2.2. Prometheus

Prometheus [27] is a methodology designed for the realisation of BDI agent systems with the use of goals and plans. It supports development activities from requirements specification through to detailed design for implementation. Prometheus design tool (PDT) [29, 30] is a graphical editor that supports the Prometheus methodology. The PDT supports the development and documentation of all the phases of the Prometheus methodology for building agent-based systems. Prometheus has three inter-connected design phases which are *system specification*, *architectural design,* and the *detailed design*.

## 3. MULTI-AGENT BASED PRE-ASSESSMENT SYSTEM DESIGN METHODOLOGY

In this section, we present the prometheus design methodology and detailed analysis of the pre-assessment system. In Table 2 are the PDT symbols and description of their functions in the design of agent-based systems. The pre-assessment system is a formative assessment system designed to supports the learning of SQL. Learning as asserted in [31] is an effort intensive task. Thus, designing a MAS for learning purposes is a complex process [28].

### 3.1. System specification

The system specification phase begins with a high-level description of the problem, which leads to the identification of *initial goals* decomposition as shown in Figure 2.

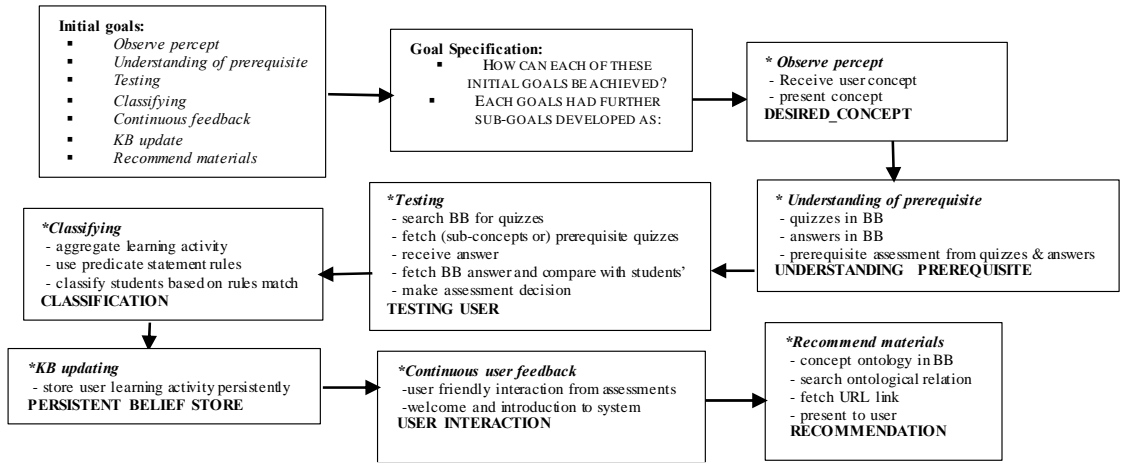

Figure 2. High level description of problem including initial goal and overall system goal specification

Table 2. The PDT notation symbol and meaning [3, 32, 33]

| Name | Symbol | Description |
| --- | --- | --- |
| Agent | Agent | The agent symbols. |
| Action | Action | This is what the agent does that has effect on the environment or other agents. |
| Role | Role | This symbolizes roles or group of roles for agents. |
| Protocol | Protocol | Protocols specifies interaction between agents. Protocols are specified using textual notations that maps to AUML2. |
| Data | Data | This is used to represent the belief (internal knowledge model) or external data. It is where functionalities that transcends to agent read or write data or information. |
| Messages | Message | This is used to symbolize a message communication between agents. |
| BDI Messages | BDIMessage | This symbol is used to represent messages that updates the beliefs of agents. |
| Percept | Percept | Represents the input coming from the environment to the agent. |
| Scenario | Scenario | This is an abstract description of a sequence of steps taken in the development of a system. It is usually the initial step that starts for the breakdown of the "statement of problem" or description of the problem to solve. |
| Goal | Goal | It is the realizable target or achievement set for an agent. |
| Connection Arrows | PDTConnection | They are edges that connects entities (i.e. symbols) together. |

#### 3.1.1. Scenario overview

Scenarios and system goals are complementary. In the process of extracting the *main goals* from the problem description, scenarios were developed as shown in Figure 3. In Figure 3 are the set of scenarios derived from the specified goals using the PDT *scenario overview* diagram.

#### 3.1.2. System *goal* diagram

The PDT system *goal overview* diagram enables the break-down or refinement of the set of derived scenarios into units of achievable design steps. The Figure 4 is the system *goal* and *subgoals* design and the interactions between them. The *AND* is a conjunction function which indicates that, at that level of design, the agent must communicate with both *classify* and the *persistentBB update* goals after its *decision-making* function. In Figure 4, at the user interface, *percept goal* is seen interacting with the *understanding of prerequisite goal* which connects to the *testing goal*. Then to the *make decision goal* that is linking both the *classify* and *persistentBB update goals* after its decision-making function; and the *classify goal* further connects the *recommend material goal*.

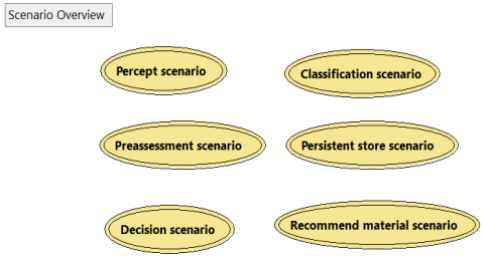

Figure 3. System scenario view

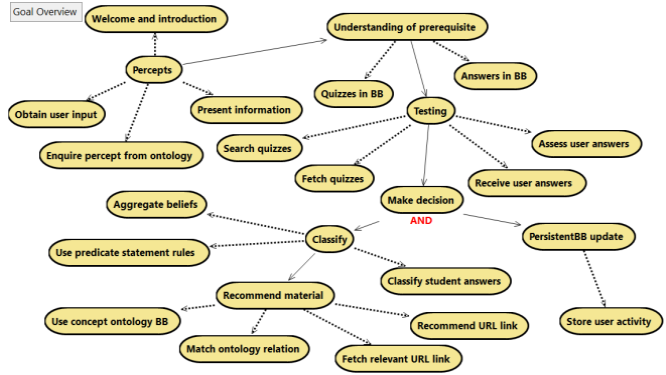

Figure 4. System goals specification and decomposition for the pre-assessment system

### 3.2. Architectural design

At this phase, the role of the system for the purpose of pre-assessment has been conceptualized. The needed number of agents and their descriptive names have been determined and included in the design. This phase covers the system overall (static) structure using *system overview* diagram, and the description of the *dynamic* behaviour of the system using interaction diagram and interaction protocols. In Figure 5 are the identified *roles* that are needed within the role decomposition results into a role hierarchy from super-roles to atomic roles (top-down direction) [28].

In this step, all the agents, their percepts, incoming messages, actions duly taken and interaction in the design are presented in Figure 6. In the *system overview diagram*, the data (agent knowledge) that is expected to used is coupled with the agents. In this design, the data are quizzes, answers to quizzes, and URL data links for each of the leafnodes (sub-topics) in the ontology. The data is modelled as internal knowledge or beliefs in the agents. Figure 6 also presents the five working agents of the system whose detail design are illustrated in *agent overview* stage.

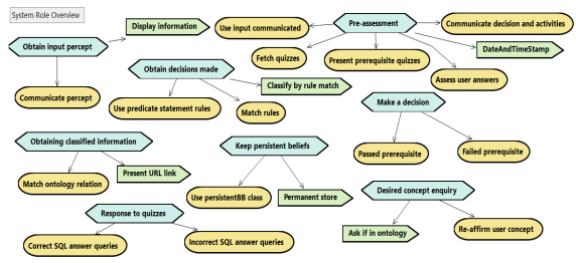

Figure 5. System role overview showing structured functionalities

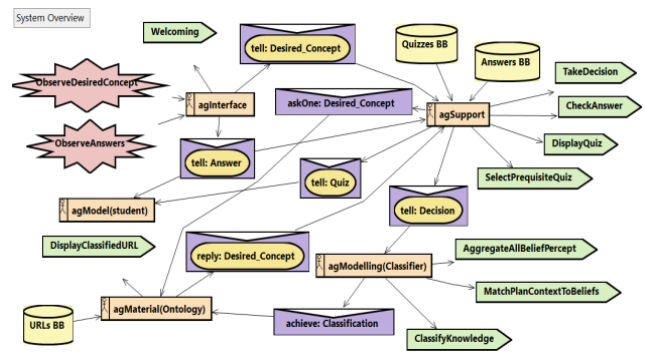

Figure 6. System overview diagram

### 3.3. Detailed design

This phase is focused on the description of responsibilities and capabilities of the internal structure of the individual agent, and how each agent would achieve their task within the system. Diagrammatically, these capabilities were realised at the *agent overview stage* shown below.

#### 3.3.1. Agent overview

In this section, the list of agents in the system and individual agent internal details are presented. This includes the agents' plan models, notation for percepts and triggering_event, communication links and inter-agent message description. At the agent overview stage, inherited interfaces from e.g., the *system* *overview* phase are adopted for specifying agents' details. The inherited interfaces are the notation symbols that appear greyish in colour.

*Agent agInterface:* In Figure 7 is a much refined and detailed design where CArtAgO artifact is the medium that was used to get user input. The interface agent first creates the artifact in order to observe it. The observed inputs are communicated as messages in agent plan (shown with the plan diagram or symbol) to other agents e.g., the agent *agSupport* that is responsible for pre-assessing students.

*Agent agSupport:* This is the pre-test agent that is saddled with the task of questioning a user's skills before making recommendation, as shown in Figures 8-9. The agent *agSupport* uses its *achievement goals* for navigation, from leafnode $z_{j,k}$ to leafnode $z_{j,k+1}$ in the hierarchy of concepts to retrieve quizzes which are represented in predicate logic in its BB to test students' skills. Using the answer percept received, it compares and matches the given answer input with the predefined answer in its BB. Taking the decision for either a *passed* or a *failed* predicate on every answer received, this agent also communicate all assessment activities, namely: the decision reached per question, the questions asked, and communication of the answers received to other agents in the MAS that needs to know. This agent also *date* and *timestamp* every learning activity.

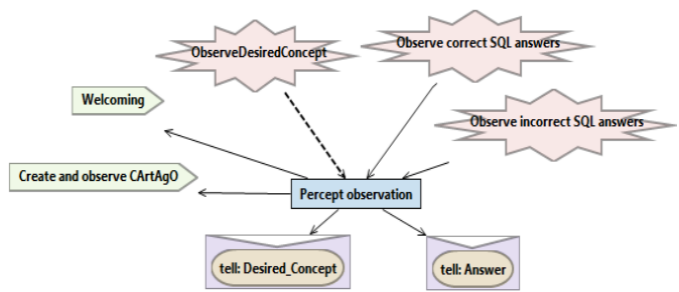

Figure 7. Detailed overview of agent agInterface

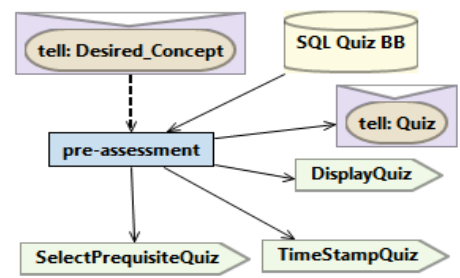

Figure 8. Agent agSupport receiving the desired_Concept percept and retrieving quizzes using plans

*Agent agModelling:* This agent gets message percepts from agent *agSupport* for every leafnode (question attached to a unit of learning) in the ontology whose pre-assessment has been completed. This agent uses the percept (or information) it receives to match the pre-conditions in its plan *context*, and thereafter classify the student's skills. The category of information (in one plan) that is determined by this agent is communicated to the next receiving agent (*agMaterial*) that will in turn send learning material to the student, as shown in Figure 10.

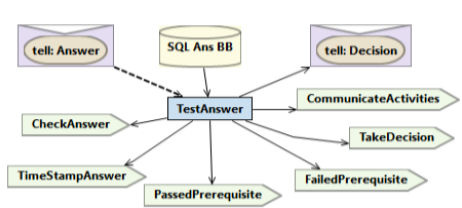

Figure 9. Agent agSupport overview

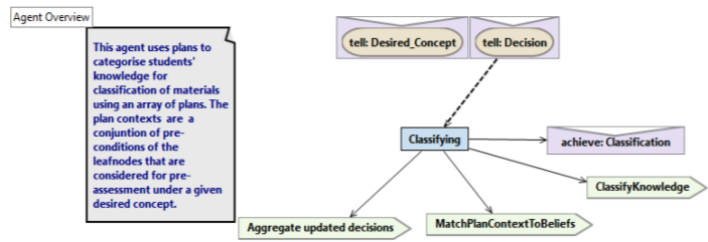

Figure 10. The agent agModelling (classifier agent)

*Agent agMaterial:* Figure 11 is agent *agMaterial* that keeps the URLs links of learning material as an ontology. The perfomative used in the message to this agent is "*achieve*". On receiving the "*achieve*" performative message from the classifier agent (after classification), the agent *agMaterial* then releases learning materials for students to learn. These materials are dependent on the number of *failed* and *passed* prerequisite assessments.

*Agent agModel:* This agent uses the Java *TextPersistentBB* class to store all the learning activities in the system. The *TextPersistentBB* class was configured in the MAS at the point of declaration or creation of the multi-agents project with the **Mas2j** [34] extension at the level of implementation. The activities stored are messages sent to the agent; and they include students' desired topics, and answers to question (both correct or incorrect) percept. As shown in Figure 12, the persistent beliefs are permanently stored in the system.

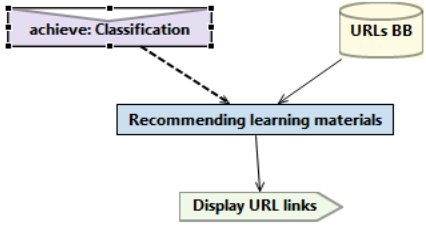

Figure 11. Agent agMaterial: The learning material agent overview

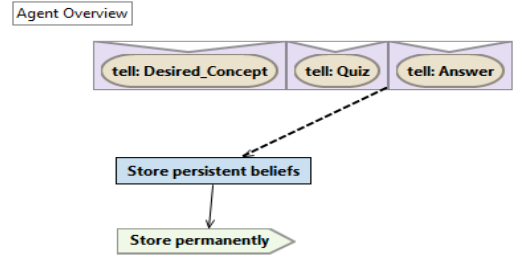

Figure 12. Agent agModel (student) overview

### 3.4. Logical analysis of pre-assessment

Previous reports on the pre-assessment system presented a pre-assessment mechanism [35, 32] and the formalized logic model [36] for pre-skills testing, decision making, selection and recommendation of learning materials. In general, on any given knowledge graph or ontology, the following symbolic algorithm presents the underlying reasoning [32]:

$\forall D\ \forall C_i\ \forall N_{i,j}\ hasPrerequisite(D_{i,j}, C_{i+1}) \land hasKB(C_{i+1,j\pm1}, N_{j\pm1,k})$
[
: $\exists D \land \forall passed(N_{j\pm1,k}) => learn(N_{j,k}) \land hasKB(D_{i,j}, N_{j,k})$
else
: $\exists D \land \exists failed(N_{j\pm1,k}) => learn(N_{j\pm1,k}) \land hasKB(C_{i+1,j\pm1}, N_{j\pm1,k})$
]

of the multi-agent pre-assessment system whose software engineering design steps we have presented in the preceding sections; such that, $D$ is the *desired concept* (also called the desired topic) that subsumes some prerequisites $C_i$ which further subsumes some leafnodes $N_{i,j}$. In description logic notation, it states $N_{i,j} \sqsubseteq C_i \sqsubseteq D$. In the system, the content of learning is in the domain of SQL (structured queried language) from which topics – that we have called the DesiredConcept $D$ are chosen and studied by students.
Now, let $D = \{\ c \in D\ |\ p(c)\ \}$ and $N = \{\ n \in C\ |\ q(n)\ \}$. $D$ precedes $C$ in the hierarchy of concepts (or topics) of learning such that the number of elements in $D = C + 1$. Then the set of topics otherwise known as elements considered in the domain $D$ is given as
  $D = \{union, join, update, delete, insert, select\}$;
and the set of all prerequisites $C$ underneath $D$ is given as
  $C = \{join, update, delete, insert, select\}$;
and the set of all terminal leafnodes $N$ in $C$ and $D$, respectively, given as
  $N = \{union, unionAll, selfJoin, fullOuterJoin, innerJoin, updateSelect, updateWhere, deleteSelect, deleteWhere, insertSelect, insertValue, selectOrderBy, selectDistinct, selectWhere, selectAll\ \}$.

In education, teaching-learning is chronological and this forms the basis for the connection of a previous learning to a new or ongoing learning. Let the relation $R$ be the set of connection between nodes i.e. a new topic and a previously learned topic. Then we state that $D$ and $C$, and $C$ and $N$ belong to some relations, respectively; as shown in the following set A and B with regards to the given ontological node relationships
  A = D x C ∈ R and, B = C x N ∈ R.
Symbolically, it holds that
  $\forall d \in D\ \forall c \in C\ \forall n \in N, R(D, C) \land R(C, N)$
In furtherance, the elements of the sets $D$, $C$ and $N$ thus satisfies the following definitions and their respective properties $p$ and $q$ in predicate formulas
  $(d, c) = \{p\ \in R\ |\ p(d, c)\}$, where the relation $p = hasPrerequisite$; and
  $(c, n) = \{q\ \in R\ |\ q(c, n)\}$, where the relation $q = hasKB$.
Symbolically, the conjunction of the above given relation is
  $\forall d \in D\ \forall c \in C\ \forall n \in N, hasPrerequisite(d, c) \land hasKB(c, n)$.

By the property of transitivity,

$hasPrerequisite(d, c) \land hasKB(c, n) \rightarrow hasKB(d, n)$.

For example, choose the *update* $\in D$. Its immediate prerequisite *delete* $\in C$. The node *update* has its leadnodes $n = \{updateSelect, updateWhere\}$, and the node *delete* also has its leafnodes $n = \{deleteSelect, deleteWhere\}$. From the foregoing,

$hasPrerequisite(update, delete) \land hasKB(delete, \{deleteSelect, deleteWhere\}) \rightarrow hasKB(update, \{deleteSelect, deleteWhere\})$.

Then, the conclusion becomes the units to be pre-tested on, and if any is failed; students are recommended materials for the failed units to learn. As stated earlier, if no pre-test is failed, i.e. all pre-assessments are passed; then the conclusion as in the following thus hold

$hasPrerequisite(update, delete) \land hasKB(delete, \{deleteSelect, deleteWhere\}) \rightarrow hasKB(update, \{updateSelect, updateWhere\})$

which are the leafnodes of the chosen topic. The stated logic-based representation, formed the ground fact internal knowledge model of the multi-agent pre-ssessment system designed in this study.

## 4. DISCUSSION

The paper has presented the prometheus AUML design tool for the design and analysis of the pre-assessment system, and its implementation with Jason – a Java-based interpreter and declarative language. The choice of Prometheus methodology ensured that every requirement and detailed design activity were captured with the appropriate symbol. This we have depicted from *initial goal specifications*, to *subgoals*, to agent *roles* and interaction using distinctive diagrams. From critical analysis, Prometheus provides support on how requirement statements may be acquired -- starting with *intial goals* specification -- as well as a general system architecture as against some other AUML tools. These steps are vital as any left-out functionality would cause a void in the system: A void that may require the re-engineering of the whole system. In a declarative language, agents communicate via message passing in predicate logic form. Thus, in line with the reported mechanism of pre-assessment & recommendation and formalized (FOL based) pre-assessment rules [35-37] in which the MAS made accurate recommendation after pre-assessment, Figure 13 hereby presents the pseudocode of the operation of the system and how the perceived knowledge by agents are used: from percept acquisition at the interface (line 7), through to other agents via the *.send() internal action* [4] (on lines 9, 11, 18, 23, 27; as shown in Figure 13) which clearly shows the number of interactive agents in the system. Between each *internal action*, is the action designated for a receiving agent to execute.

```
1.  initial beliefs: predicate(Class, Class)
2.  initial beliefs: predicate(Class, Leafnode)
3.  initial beliefs: predicate(Leafnode, URL)
4.  initial beliefs: quiz(PrerequisiteLeafnode)
5.  Given a desired concept that has N leafnodes prerequisite
6.  IF
7.     Percept ← desiredConcept
8.  THEN
9.     .send(receiver, tell, desiredConcept)
10.    fetch the next quiz(Prerequisite_Leafnode)
11.    .send(receiver, tell, quiz(Prerequisite_Leafnode)
12.    output quiz(Prerequisite_Leafnode)
13.    Percept ← answer(X)
14. IF
15.    answer(X) == answer(Prerequisite_Leafnode)
16. THEN
17.       passed(Prerequisite_Leafnode) decision
18.       .send(receiver, tell, passed(Prerequisite_Leafnode)
19.    IF
20.       answer(X) \== answer(Prerequisite_Leafnode)
21.    THEN
22.       failed(Prerequisite_Leafnode) decision
23.       .send(receiver, tell, failed(Prerequisite_Leafnode)
24. IF
25.    N number of leafnodes have been pre-assessed on
26. THEN
27.    .send(receiver, achieve, recommendMaterial)
28. Else
29.    repeat 10 to 27
```

Figure 13. Pseudo-algorithm of the pre-assessment process that depends on the number of leafnodes N considered under a desiredConcept

**4.1. Resolving issues of development**

As a pre-assessment system for SQL pre-skills test and skills classification into two [1, 0] binary states, the system receives *open-ended* SQL query inputs that maybe correct or incorrect answer for a particular query question. For a given query, the inflow of incorrect answers to the system is not definite nor predetermined compared to correct query answers that are known and predefined in the system. Thus, programming a MAS for the recognition of *negative facts* (i.e. incorrect answers) can pose some difficulty for agent plan selection and execution of agent *goals* when the expected inputs may vary depending on a student understanding and query competences. In such cases, inputs become diverse, unbounded and subjective to the students. On the hand, the correct SQL queries which are the *positive facts* are quite straightforward to program because every answer to the questions asked on the system is predetermined based on standard SQL queries. Typically, the syntax of Jason agent plan comprises three parts, and the structure given as: *triggering_event, context <-- body* [4]. The *context* is the part of the plan which states the pre-condition that activates a plan for execution. By default, a blank plan *context* is true for all beliefs in the agent. Otherwise, a predicate form pre-condition must be stated to control what plan is right for a given *triggering_event* and beliefs. To determine whether a query input is right or wrong, the predefined *positive facts* were represented in the pre-assessment agent in first order logic (FOL) predicate form – for a declarative language.

When an agent gets percept (also in FOL form), the agent matches that percept (now a belief) against all plans, the relevant plan is selected and actions in the *body* of the *plan* is executed --- in *One vs. All* approach. This is because *positive facts* are information whose representation are known and can be represented or given to the agent (i.e. responsible for handling pre-skills assessment) for comparison with incoming percepts. So, the correct SQL queries were initialized in the agent's belief base and were used by the agent to match and trigger relevant plans and agent *goals* as needed. But *negative facts* are unknown and as such cannot be pre-determined for representation as mentioned earlier. So, to address incorrect SQL query inputs, Jason *different* \== operator [4] was used as the comparison operator in the agent plan *context*. In analogy, the operator means "**not equal**" or "**false**". But the use of this operator was not without inconsistency in the collective multi-agents behaviour during system coding and implementation phase. During coding, when an incorrect query was inputted for testing, the \== operator made the pre-assessment agent to miss-select plans from its plan library. For example, by stating in a plan that was expected to handle an incorrect SQL query, that,

> *if the answer percept coming into the system **does not** match some initially predefined SQL queries **then** inform the student that the answer given is incorrect and **then** select the next leafnode question and present to the student.*

Literally, from the behaviour exhibited by the agent, the agent's interpretation was any other plan whose plan *context* has no match to any already known knowledge in the agent's belief base. The miss-selection of plans was due to some uncertainty in the agent ability to map an incorrect query percept to beliefs. This behaviour as observed adversely altered the order of subsequent goal/question selection of a prerequisite's leafnode *N*, in contrast to the arrangements of nodes in the ontology tree. This was a non-trivial problem. At the implementation phase, one of the key principles of software methodology is to combine *coding* and *testing* [38]. This principle which enables a system to be investigated while it is still being developed ensured that this non-trivial problem was checked before the system was completely built.

To enable the pre-assessment agent, as shown in Figures 8-9 to accurately select relevant plan(s) for a match of its plan context to the percept that is adopted in the \== operator; and to correctly determine the next appropriate agent goal and accurate message passing to other agents, we had to introduce a process of iteration that could count plan selection in the agent for every parent node (or topic) and their connected leafnodes *N*. In Jason predicate logic form, an example syntax of this iteration is *countForDeletePre*(*X*), which depicts the counter for the *delete* node where X is a positive integer. In addition, the negation of some incoming percept was required to stop unsolicited plan trigger. An example of such negation in the *context* part of a plan, was the *not desiredConcept*("*insert*") which was used to block-off the *desiredConcept*("*insert*") in a plan *context* so as not trigger the wrong plan and wrong agent goal at a given time. As the system kept expanding with the programming of more parent nodes *D* and leafnodes *N* been added, this block-off continued and it used to mitigate agent anomaly behaviour. These two combined strategies effectively controlled agent behaviour as well as the entire multi-agents toward handling of the incorrect SQL query inputs.

### 4.2. Experimentation and results

SQL programming has been reported to pose challenges to student learners e.g., [39]. Our previous survey conducted in [32] also found that students are faced with SQL query difficulties. In this paper, we have slightly varied the approach to prior-skills assessment that was taken in previous survey. This was to determine whether changing any *factor* before and during the course of pre-assessment can cause any significant change in the results obtained in previous researches. Against previous research, the two approaches introduced are, namely: *i) To allow students have adequate familiarity with the database (which has five tables) prior to pre-skill assessments*; and *2) To allow students consult with any text or written materials throughout the time of their pre-skill assessments on the system*. But with no prior access to the questions attempted at their pre-skills evaluation. From the results obtained in this paper, changes in the approach of pre-testing has slightly improved results, as shown in Tables 3-4 on student pre-skills exercise in contrast previous surveys in [32, 39, 40]. As students used the system, the system stored all pre-assessment activities and also *timestamped* every entry made by students into the system. Figure 14 is a screenshot of the pre-assessment system interface that is presenting the ***Union*** desired topic, and subsequent interaction between the student and the MAS in the course of pre-assessment. Agents have independent autonomy and mental states. The MAS leveraged on these properties coupled with their interaction protocol [7] and we obtained optimal recommendation of learning materials for either all *passed* or any *failed* pre-assessments.

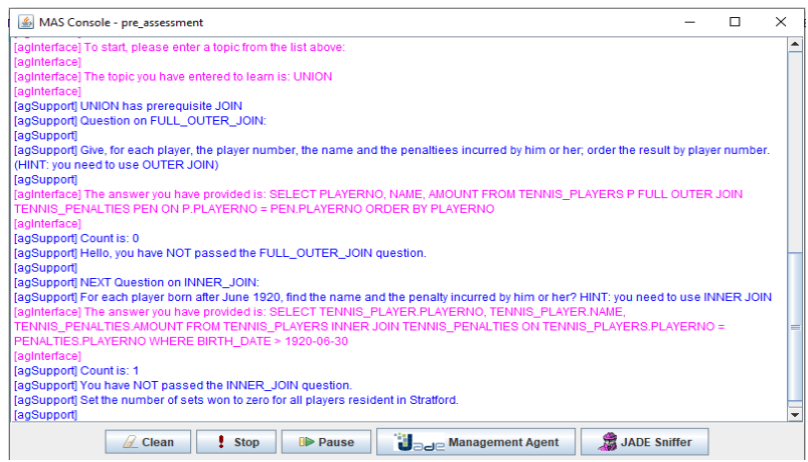

Figure 14. Agent interface showing the Union desired concept

#### 4.2.1. Agent mind inspection

Upon the agent *agModel Mind Inspection*, some examples of the structured beliefs stored by this *PersistentBB* agent in FOL predicate form are:

- Beliefs  desired_Concept("UNION, date(2020-3-23),time(14-4-20)")[source(agSupport)]

which is the belief about a student's desired topic, and

responseToFullOuterJoin("SELECT PLAYERNO, NAME, AMOUNT FROM TENNIS_PLAYERS P FULL OUTER JOIN TENNIS_PENALTIES PEN ON P.PLAYERNO = PEN.PLAYERNO ORDER BY PLAYERNO, date(2020-3-23), time(14-17-8)")[source(agSupport)]

the persistence belief of the student's response to a pre-assessment on *Full Outer Join* query; and

responseToInnerJoin("SELECT TENNIS_PLAYER.PLAYERNO, TENNIS_PLAYER.NAME, TENNIS_PENALTIES.AMOUNT FROM TENNIS_PLAYERS INNER JOIN TENNIS_PENALTIES ON TENNIS_PLAYERS.PLAYERNO = PENALTIES.PLAYERNO WHERE BIRTH_DATE > 1920-06-30, date(2020-3-23), time(14-23-21)")[source(agSupport)]

the persistence belief of a response to a pre-assessment on *Inner Join* query*; and then*

failed("The student has NOT passed the INNER_JOIN question., date(2020-3-23), time(14-23-21)")[source(agSupport)]
failed("The student has NOT passed the FULL_OUTER_JOIN question., date(2020-3-23), time(14-17-8)")[source(agSupport)]

the persistence belief of *failed(N)* pre-assessments on *Full Outer Join* and *Inner Join.*

In Table 3 is the data and the outcomes of either *passed*(*N*) or *failed*(*N*) binary states [1, 0] for each leafnode *N* and the timespent on each leafnode *N* pre-assessment task. Recall, Use the "Insert Citation" button to add citations to this document.
Use the "Insert Citation" button to add citations to this document. In [33] earlier that we gave the set of leafnodes *N* as

> *N = {union, unionAll, selfJoin, fullOuterJoin, innerJoin, updateSelect, updateWhere, deleteSelect, deleteWhere, insertSelect, insertValue, selectOrderBy, selectDistinct, selectWhere, selectAll}.*

Now, for the purpose of statistical plots and presentation, let *N* be encoded a set of positive integers such that we have *N = {1, 2, 3, 4, 5, 6, 7, 8, 9, 10, 11, 12, 13, 14, 15}*
where *N* = 1,2, 3, …, 15 are nominal values. Table 3 presents the encoded leafnode *N* information and the total number of attempted pre-skill tests. In Table 4 are the *passed*(*N*) or *failed*(*N*) per the time spent on each leafnode *N*, respectively. From Table 3, it is observed that the changes introduced has caused some significant changes in the results as against our previous findings.

Table 3. Leafnode encoding and number of Passed(N) and Failed(N)

| Leafnodes N | Leafnode Encoding | No. of Passed(N) Pre-assessment | No. of Failed(N) Pre-assessment |
|---|---|---|---|
| Union [U] | 1 | - | - |
| unionAll [UA] | 2 | - | - |
| selfJoin [SJ] | 3 | 11 | 2 |
| fullOuterJoin [FOJ] | 4 | 6 | 7 |
| innerJoin [IJ] | 5 | 4 | 9 |
| UpdateSelect [US] | 6 | 1 | 15 |
| updateWhere [DW] | 7 | 16 | 0 |
| deleteSelect [DS] | 8 | 5 | 16 |
| deleteWhere [DW] | 9 | 20 | 1 |
| insertSelect [IS] | 10 | 3 | 17 |
| insertInto [II] | 11 | 16 | 4 |
| selectOrderBy [SOB] | 12 | 11 | 3 |
| selectDistinct [SD] | 13 | 14 | 1 |
| selectWhere SW] | 14 | 14 | 0 |
| selectAll [SA] | 15 | 14 | 0 |

Visualization of the data is presented in Figures 15-17. The data was plotted using 80% training and 20% test. Figure 15 shows the scatter plot of the *timespent* against the leafnode *N* encoded as integer values with the display of the respective *leafnode N* per *timespent*. In Figure 16 is the scatter plot of linear regression model. From the plot, the linear model predicts that there would be an increase *passed(N)* ≡ 1 binary state as the *timespent* on pre-assessment tasks decreases. Invariably in the plot, there is a correlation between increase in *passed pre-assessments* and *continuous decrease in time spent*; which implies increase in the recommendation of chosen desired topics. Figure 17 is a plot of the logistic regression model. Like the linear regression model, the model also predict increase in *passed(N)* pre-assessments. That is, in future more students are likely to pass their pre-assessments in the domain of SQL, if and only if, the two varied approaches introduced here are kept and adopted.

Table 4. Boolean classification [1, 0] and time spent on each pre-assessment task *(continue)*

| Leafnode Encoding | Boolean Classification [1, 0] vs. Timespent (mm:ss) |
|---|---|
| 1 | Nil \| Nil |
| 2 | Nil \| Nil |
| 3 | [1] \| (12:09), [1] \| (09:11), [1] \| (05:33), [1] \| (10.01), [1] \| (04:21), [1] \| (11:12), [1] \| (05:21), [1] \| (08: 01), [1] \| (04: 07), [1] \| (07:25), [1] \| (08: 45), [0] \| (10:12), [0] \| (07:13) |
| 4 | [1] \| (05:16), [1] \| (13:02), [1] \| (10:22), [1] \| (06: 56), [1] \| (11:34), [1] \| (15:08), [0] \| (9:19), [0] \| (5:33), [0] \| (16:48), [0] \| (17:59), [0] \| (06:41), [0] \| (05: 00), [0] \| (11:54) |
| 5 | [1] \| (05: 55), [1] \| (04:35), [1] \| (16: 24), [1] \| (07: 31); [1] \| (02:47), [1] \| (06:57), [0] \| (09:35), [0] \| (09:12), [0] \| (11:43), [0] \| (05:13), [0] \| (11:48), [0] \| (13:10), [0] \| (14: 19) |
| 6 | [1] \| (19:00), [0] \| (20:03), [0] \| (13: 44), [0] \| (07: 11), [0] \| (15:17), [0] \| (15: 08), [0] \| (03:51), [0] \| (08: 10), [0] \| (02: 14), [0] \| (01: 46), [0] \| (15: 16), [0] \| (18:05), [0] \| (11: 10), [0] \| (03: 49), [0] \| (14:10), [0] \| (09:43) |
| 7 | [1] \| (01:23), [1] \| (01: 58), [1] \| (04: 11), [1] \| (11:29), [1] \| (03:14), [1] \| (15:10), [1] \| (11.21), [1] \| (08:41), [1] \| (11:03), [1] \| (05:51), [1] \| (15: 09), [1] \| (04: 17), [1] \| (04:16), [1] \| (01: 44), [1] \| (03:17), [1] \| (11: 04) |
| 8 | [1] \| (10: 26), [1] \| (12:05), [1] \| (13: 02), [1] \| (17: 33); [1] \| (12:24), [0] \| (11:15), [0] \| (03: 45), [0] \| (07: 30), [0] \| (11:19), [0] \| (05: 18), [0] \| (03:55), [0] \| (18: 00), [0] \| (03: 44), [0] \| (21: 40), [0] \| (07: 25), [0] \| (11:37), [0] \| (19: 16), [0] \| (02: 41), [0] \| (14:12), [0] \| (08:13), [0] \| (04: 58) |

Table 4. Boolean classification [1, 0] and time spent on each pre-assessment task

| Leafnode Encoding | Boolean Classification [1, 0] vs. Timespent (mm:ss) |
|---|---|
| 9 | [1] | (02:22), [1] | (10: 15), [1] | (06: 08), [1] | (06:11), [1] | (05:40), [1] | (02:15), [1] | (01.58), [1] | (02:12), [1] | (17:22), [1] | (09:21), [1] | (08: 39), [1] | (07: 47), [1] | (07:15), [1] | (01: 16), [1] | (12:18), [1] | (01: 54), [1] | (15:15), [1] | (07: 11), [1] | (11:18), [1] | (11: 54), [0] | (01:32) |
| 10 | [1] | (02:22), [1] | (01:23), [1] | (07: 36), [0] | (01:55), [0] | (01: 57), [0] | (04:10), [0] | (11:31), [0] | (02:20), [0] | (14:25), [0] | (03: 00), [0] | (03: 58), [0] | (08:14), [0] | (06: 47), [0] | (12:37), [0] | (05: 21), [0] | (12:17), [0] | (11:12), [0] | (04:11), [0] | (07:15), [0] | (08:18) |
| 11 | [1] | (00:59), [1] | (06: 50), [1] | (02: 01), [1] | (01:29), [1] | (01:45), [1] | (03:04), [1] | (03:22), [1] | (04:01), [1] | (07:23), [1] | (05:11), [1] | (02:48), [1] | (04:07), [1] | (04:10), [1] | (09: 10), [1] | (05:31), [1] | (01:23), [0] | (02:59), [0] | (02:21), [0] | (05: 26), [0] | (13:05) |
| 12 | [1] | (03: 56), [1] | (04:00), [1] | (00:53), [1] | (07:34); [1] | (01:19), [1] | (03:12), [1] | (06:22), [1] | (04:31), [1] | (05: 12), [1] | (07:04), [1] | (07:17), [0] | (01: 40), [0] | (04:51), [0] | (06:33) |
| 13 | [1] | (01:01), [1] | (01: 51), [1] | (02: 28), [1] | (03:12), [1] | (03:41), [1] | (02:35), [1] | (07:48), [1] | (03:27), [1] | (07:16), [1] | (04:43), [1] | (01: 59), [1] | (02: 55), [1] | (04: 17), [1] | (03: 26), [0] | (02: 38) |
| 14 | [1] | (03:21), [1] | (11: 14), [1] | (00: 56), [1] | (05:21), [1] | (04:11), [1] | (02:22), [1] | (00:50), [1] | (02:32), [1] | (00:26), [1] | (03:39), [1] | (02: 01), [1] | (04: 21), [1] | (04: 15), [1] | (04: 44) |
| 15 | [1] | (00:37), [1] | (02: 55), [1] | (00: 28), [1] | (01:58), [1] | (03:01), [1] | (01:45), [1] | (01:13), [1] | (01:42), [1] | (05:29), [1] | (01:17), [1] | (01: 30), [1] | (04: 27), [1] | (00:23), [1] | (00: 34) |

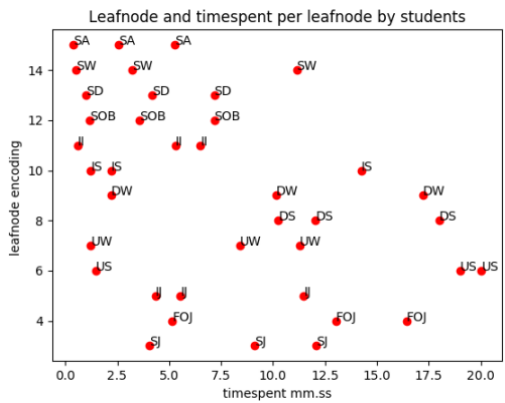

Figure 15. Scatter plot of timespent per leafnode $N$

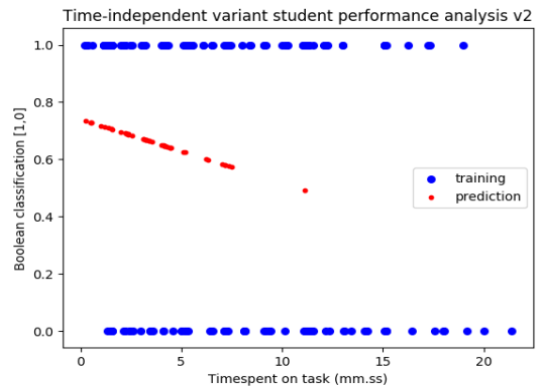

Figure 16. Linear regression plot of timespent to boolean classification

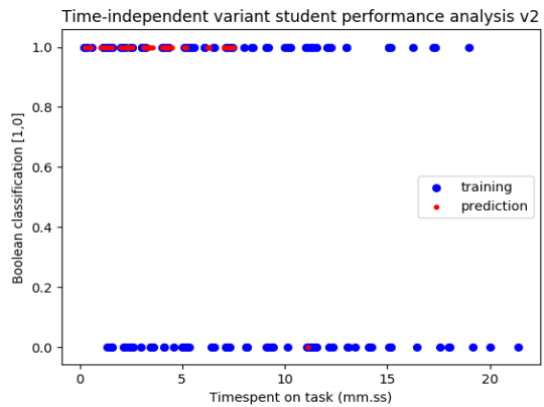

Figure 17. Logistic regression plot on timespent to boolean classification

## 5. CONCLUSIONS

This paper has presented a detailed analysis and design of a formative elearning multi-agent pre-assessment system using the Prometheus methodology design tool and implementation with Jason (Agentspeak) language. Detailed description and functions of the agent has been presented using different diagrams of the respective agents and their roles. The paper covered all design activities including issues that evolved during implementation and the solution strategy that was adopted; then to evaluation, data collection and analysis of data. The design activity also covered percept observation by the interface agent, to inter-agent communication, decision making strategy, classification of user skills and recommendation of materials for students' study. The content of the system is databases/SQL: a subject that has been asserted to have pose difficulty to students. This project was designed to identify the gaps between what a student wants to learn and what the student has already learned. The two conditions introduced during the pre-skill tests in this paper have shown that changes in certain factors can change the narratives of the difficulty faced by students in SQL programming. Further work is the formalization of agent rules using a formalized language in the pre-assessment and recommendation strategy.

**REFERENCES**
[1] UML, "Introduction to OMG's Unified Modeling Language," 5 Feb 2020. [Online]. Available:http://www.omg.org/ gettingstarted/what_is_uml.htm.
[2] S. Stoyanov, I. Ganchev, I. Popchev and M. O'Droma, "An Approach for the Development of InfoStation-Based eLearning Architectures," *Compt. Rend. Acad. Bulg. Sci*, vol. 61, pp. 1189-1198, 2008.
[3] RMIT, "Agents Group," RMIT University, Australia, 2012. [Online]. Available: https://sites.google.com/site/rmitagents/software/prometheusPDT/tutorials. [Accessed 5 Apr 2020].
[4] R. H. Bordini, J. F. Hübner and M. Wooldridge, Programming s in AgentSpeak using jason, John Wiley & Sons, 2007.
[5] C. Rouveirol and V. Ventos, "Towards learning in CARIN-ALN," in *In International Conference on Inductive Logic Programming*, Berlin, Heidelberg, 2000.
[6] S. Konstantopoulos and A. Charalambidis, "Formulating description logic learning as an inductive logic programming task.," in *In International Conference on Fuzzy Systems*, 2010.
[7] J. Chen and J. Li, "Globally fuzzy leader-follower consensus of mixed-order nonlinear s with partially unknown direction control," *Information Sciences,*, vol. 523, pp. 184-196, 2020.
[8] W. A. Munassar and A. F. Ali, "Semantic Web Technology and Ontology for E-Learning Environment," *Egyptian Computer Science Journal,* vol. 43, no. 2, pp. 88-100, 2019.
[9] A. S. Aziz, S. A. Taie and R. A. El-Khoribi, "The Relation between the Learner Characteristics and Adaptation Techniques in the Adaptive E-Learning Systems.," in *International Conference on Innovative Trends in Communication and Computer Engineering (ITCE)*, 2020.
[10] A. Trifa, A. Hedhili and W. L. Chaari, "Knowledge tracing with an intelligent agent, in an e-learning platform," *Education and Information Technologies,* vol. 24, no. 1, pp. 711-741, 2019.
[11] U. C. Apoki, S. Ennouamani, H. K. M. Al-Chalabi and G. C. Crisan, "A Model of a Weighted Agent System for Personalised E-Learning Curriculum," in *International Conference on Modelling and Development of Intelligent Systems*, 2019.
[12] F. El Hajj, A. El Hajj and R. A. Chehade, "Vulnerability detector for a secured E-learning environment," in *Sixth International Conference on Digital Information Processing and Communications (ICDIPC)*, Beirut, 2016.
[13] C. Anghel and I. Salomie, "JADE Based solutions for knowledge assessment in eLearning Environments," *EXP-in search of innovation (Special Issue on JADE),* 2003.
[14] N. Stancheva, I. Popchev, A. Stoyanova-Doycheva and S. Stoyanov, "Automatic generation of test questions by software agents using ontologies," in *In 8th International Conference on Intelligent Systems (IS)*, 2016.
[15] V. M. F. Ferreira, J. C. C. Carvalho, R. M. E. M. da Costa and V. M. B. Werneck, "Developing an educational medical game using AgilePASSI multi-agent methodology," in *In 28th International Symposium on Computer-Based Medical Systems (CBMS),*, 2015.
[16] X. Lan, L. Liu and Y. Wang, "ADP-Based Intelligent Decentralized Control for s Moving in Obstacle Environment.," *IEEE Acess,* vol. 7, pp. 59624-59630, 2019.
[17] G. Al-Hudhud, "Designing e-Coordinator for improved teams collaboration in graduation projects," *Computers in Human Behavior,* no. 15, pp. 640-644, 2015.
[18] M. Wooldridge, N. Jennings and D. Kinny, "The Gaia methodology for agent-oriented analysis and design," *Autonomous Agents and s,* vol. 3, no. 3, pp. 285-312, 2000.
[19] L. Cernuzzi and F. Zambonelli, "Gaia4E: A Tool Supporting the Design of MAS using Gaia," in *In ICEIS (4)*, 2009.
[20] N. R. Jennings and M. J. Wooldridge, "Applications of intelligent agents," 1998.
[21] T. I. Zhang, E. Kendall and H. Jiang, "An agent-oriented software engineering methodology with application of information gathering systems for LCC," in *In for LLC, Procs AOIS-2002.*, 2002.
[22] P. Bresciani, A. Perini, P. Giorgini, F. Giunchiglia and J. Mylopoulos, "Tropos: An agent-oriented software development methodology," *Autonomous Agents and s,* pp. 8(3), 203-236, 2004.
[23] M. Morandini, D. C. Nguyen, L. Penserini, A. Perini and A. Susi, "Tropos Modeling, Code Generation and Testing with the Taom4E Tool," in *In iStar*, 2011.
[24] S. A. DeLoach, "Analysis and Design using MaSE and agentTool," Air force inst of tech wright-patterson afb oh school of engineering and management., 2001.
[25] M. Cossentino and C. Potts, "A CASE tool supported methodology for the design of s," in *In International Conference on Software Engineering Research and Practice (SERP'02)*, 2002.
[26] M. Cossentino, "From requirements to code with the PASSI methodology," *Agent-oriented methodologies, 3690,* pp. 79-106, 2005.
[27] L. Padgham and M. Winikoff, "Developing intelligent agent systems: A practical guide," 2004.

[28] S. J. Juneidi and G. A. Vouros, "Agent role locking (ARL): theory for multi agent system with e-learning case study.," in *IADIS AC*, 2005.
[29] L. Padgham, J. Thangarajah and M. Winikoff, "Prometheus Design Tool," in *In AAAI*, 2008.
[30] Z. Zhang, J. Thangarajah and L. Padgham, "Automated unit testing intelligent agents in PDT," in *In Proceedings of the 7th international joint conference on Autonomous agents and multiagent systems: demo papers*, 2008.
[31] N. Manouselis, H. Drachsler, R. Vuorikari, H. Hummel and R. Koper, "Recommender systems in technology enhanced learning.," *Recommender systems handbook,* pp. 387-415, 2011.
[32] K. E. Ehimwenma, A multi-agent approach to adaptive learning using a structured ontology classification system, Sheffield UK: Doctoral thesis, Sheffield Hallam University, 2017.
[33] "AUML-2 & Interaction Diagram Tool," [Online]. Available: http://waitaki.otago.ac.nz/~michael/auml/.
[34] R. H. Bordini, J. F. Hübner and D. M. Tralamazza, "Using Jason to implement a team of gold miners," in *In International Workshop on Computational Logic in s*, Berlin Heidelberg, 2006.
[35] K. Ehimwenma, M. Beer and P. Crowther, "Pre-assessment and learning recommendation mechanism for a Multi-agent System," in *In 14th International Conference on Advanced Learning Technologies (ICALT)*, Sheffield UK, 2014.
[36] K. E. Ehimwenma, P. Crowther and M. Beer, "Formalizing logic based rules for skills classification and recommendation of learning materials," *International Journal Information Technology and Computer Science (IJITCS),,* vol. 10, no. 9, pp. 1-12, 2018.
[37] K. E. Ehimwenma, M. Beer and P. Crowther, "Computational Estimate Visualisation and Evaluation of Agent Classified Rules Learning System," *International Journal of Emerging Technologies in Learning (IJET),* vol. 11, no. 1, pp. 38-47, 2016.
[38] A. Dennis, B. H. Wixom and D. Tegarden, Systems Analysis and Design with UML, Wiley, 2009.
[39] J. C. Prior, "Online assessment of SQL query formulation skills," *In Proceedings of the fifth Australasian conference on Computing education, Australian Computer Society, Inc,* vol. 20, pp. 247-256, 2004.
[40] K. E. Ehimwenma, M. Beer and P. Crowther, "Student Modelling and Classification Rules Learning for Educational Resource Prediction in a Multiagent System," in *7th Computer Science and Electronic Engineering Conference (CEEC2015)*, 2015.

## BIOGRAPHIES OF AUTHORS

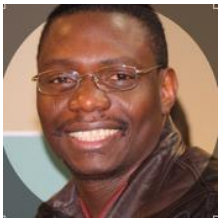

**Kennedy E. Ehimwenma** has his PhD in s, knowledge representation and rule-based logic in the area of eLearning application development at the Sheffield Hallam University, United Kingdom. His research interests include intelligent agent learning, semantic ontology, rule-based logic, and decision support systems. Dr. Ehimwenma is a lecturer in the Department of Computer Science, Wenzhou-Kean University. Email: kehimwen@kean.ed, letters4ken@gmail.com. ORCID: https://orcid.org/0000-0002-7616-9342

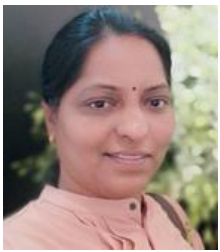

**Sujatha krishnamoorthy** is an Assistant professor of Computer Science at Wenzhou-kean University. Dr. Krishnamoorthy received her Ph. D in Information and Communication Technology, and her area of research covers digital image processing with image fusion, machine learning and computer vision. Email: sujatha@wku.edu.cn, sujatha.ssps@gmail.com [01:28, 04/05/2020] Dr. Sujatha: orcid:0000-0002-0122-6357